\documentclass[showpacs,twocolumn,preprintnumbers,showkeys,superscriptaddress,amsmath,amssymb,nofootinbib]{revtex4}
\usepackage{amsmath}
\usepackage{latexsym}
\usepackage{amssymb}
\usepackage{pstricks,pst-coil}
\usepackage{amsmath}
\usepackage{amssymb}
\usepackage[utf8]{inputenc}
\usepackage{amsfonts}
\newcommand{\be}{\begin{equation}}
\newcommand{\ee}{\end{equation}}
\newcommand{\ba}{\begin{eqnarray}}
\newcommand{\ea}{\end{eqnarray}}
\newcommand{\p}{\partial}
\def\ni{\noindent}

\begin{document}


\title{Classical electrodynamics and gauge symmetry of the $X$-boson}

%
\author{Mario J. Neves} \email{mariojr@ufrrj.br}
\affiliation{Grupo de F\'isica Te\'orica e
F\'isica Matem\'atica, Departamento de F\'isica, Universidade Federal Rural do
Rio de Janeiro, BR 465-07, 23890-971 Serop\'edica, Rio de Janeiro, Brazil}

\author{Lucas Labre} \email{lucaslabre@cbpf.br}
\affiliation{Centro Brasileiro de Pesquisas F\'isicas, Rua Xavier Sigaud, 150, 22290-180, Rio de Janeiro, RJ, Brazil}

\author{Everton M. C. Abreu}
\email{evertonabreu@ufrrj.br}
\affiliation{Grupo de F\'isica Te\'orica e
F\'isica Matem\'atica, Departamento de F\'isica, Universidade Federal Rural do
Rio de Janeiro, BR 465-07, 23890-971 Serop\'edica, Rio de Janeiro, Brazil}
\affiliation{Departamento de
F\'isica, Universidade Federal de Juiz de Fora, 36036-330, Juiz de
Fora, MG, Brazil}








\date{\today}





\begin{abstract}
\ni 
In this paper we have obtained several new results concerning the $X$-boson, which is being considered recently as one of the main candidate of the dark matter particle content.
The classical electrodynamics for the $X$-boson model was explored to understand its propagation in space-time.
The field equations (Maxwell-type) and the corresponding wave equations were obtained.
They indicate the dispersion relations of both a massive and massless particles.   These results can be interpreted as a photon and the
$X$-boson (with a mass of $17 \, \mbox{MeV}$), respectively.
The gauge symmetry and gauge transformations were discussed.
A full model diagonalization was introduced to obtain a Maxwell sector added up to a Proca sector.
After that, we have obtained the retarded Green functions which yield the inhomogeneous solutions of the wave equations
for the $X$-boson fields. We have applied this solution to the charge point motion, and we transformed the potential solution to a quadrature problem.
In particular, as an example, we have analyzed the point charge in a straight line uniform motion, where the corresponding electromagnetic
charge were calculated.

\end{abstract}



\maketitle



\section{Introduction}
The standard knowledge of the fundamental interactions tells us that there are four known forces in nature.
The particles responsible for their mediation are the photon, the $W^{\pm}$ and $Z$ bosons, the gluon and the graviton.
If a fifth force exists, it must be either short handed or it weakly interacts with the Standard Model content.
Or it can be both, if we want a consistency with the enormous quantity of experimental data.
The heightened interest in this fifth force has been reinforced by the obvious necessity for dark matter.
This last one, in its turn, has motivated new particles and forces in a hidden or dark sector
that ``mingles" with the visible one. This mixture induces a weak fifth force between the known particles.

Nuclear decays have revealed a possible existence of extra bosons as being a solution for experimental anomalies.
Recently, an anomaly of $6.8\,\sigma$ in the decay of excited Beryllium-$8$ into its ground state has
suggested the presence of a new particle in the scenario, called $X$-boson, which has mass around the $17 \, \mbox{MeV}$, and in the large pair correlation angle-near $140^o$-region. This observation attracted the attention of the theoretical scientific community since the explanation of this anomaly can originate new particles, see \cite{AJ,FengPRL2017,JFengPRD2017,GuHe2016,JiaEPJC2016,KitaharaPRD2017,EllwangerJHEP2016,Chen2017,SetoPRD2017,KozaczukPRD2017,Krasnikov2017,MJNeves2018}.
This decay process can be shown by $8 \mbox{Be}^{\ast} \, \rightarrow \, 8 \mbox{Be} + X$, where $X$ immediately decays into the electron-positron pair, {\it i. e.}, $X \, \rightarrow \, e^{+} \, e^{-}$. The $X$-boson must be a spin-$1$
Abelian field that mixes {\it kineticaly} with the electromagnetic (EM) photon.

The model representing the $X$-boson dynamics is set out by the effective Lagrangian \cite{JFengPRD2017}
\begin{equation}\label{Lagrangiana}
{\cal L} = -\dfrac{1}{4} F_{\mu\nu}^2 - \dfrac{1}{4} X_{\mu\nu}^2 + \dfrac{\chi}{2} X_{\mu\nu}F^{\mu\nu} + \dfrac{1}{2} \, m_X^2 \, X_{\mu}^2
- K_{\mu}X^{\mu} \; .
\end{equation}
Here, $F^{\mu\nu}$ and $X^{\mu\nu}$ are the Maxwell and $X$-boson strength field tensors respectively, and as usual, it is defined by $F^{\mu\nu}=\partial^{\mu}A^{\nu}-\partial^{\nu}A^{\mu}$ and $X^{\mu\nu}=\partial^{\mu}X^{\nu}-\partial^{\nu}X^{\mu}$. The $A^{\mu}$ and $X^{\mu}$ sets out the electromagnetic photon and $X$-boson fields, respectively. The $\chi$-parameter is a dimensionless kinetic mixing estimated at $10^{-6}<\chi<10^{-3}$. The $X$-mass in (\ref{Lagrangiana}) is adopted by the value of $m_{X}=17 \, \mbox{MeV}$, and it spoils the gauge symmetry under $X^{\mu}$-field. In this paper we will obtain a dual gauge invariant action for the $X$-boson.

The source $K^{\mu}$ is the current associated with the interaction between both
the Standard Model (SM) fermions with the $X$-boson, and it is given by
\begin{eqnarray}\label{Jcurrent}
K^{\mu}=\!\!\!\!\sum_{\Psi \, = \, e \, , \, \nu \, , \, u \, , \, d \, ,...} e \, \chi_{\Psi} \, \bar{\Psi} \, \gamma^{\mu} \, \Psi \; ,
\end{eqnarray}
where $\Psi$ is any fermion of the SM, and the $\chi_{\Psi}$-parameters are estimated by \cite{FengPRL2017}
\begin{eqnarray}\label{chiconstraints}
2 \times 10^{-4} < \!&|\chi_{e}|&\! < 1.4 \times 10^{-3} \; ,
\nonumber \\
|\chi_{n}| \!& < &\! 2.5 \times 10^{-2} \; ,
\nonumber \\
 \sqrt{|\chi_{\nu} \, \chi_{e}|} \!& \lesssim &\! 7 \times 10^{-5} \; .
\end{eqnarray}
In this approach, the current is a vectorial one, but there is also the approach in which the current can be axial, see \cite{KozaczukPRD2017,KahnJHEP2017}. Therefore, the $X$-boson can emerge as a new fundamental interaction in nature \cite{GuHe2016}. Some recent papers introduce the unification of $X$-boson with the others mediators bosons of the SM \cite{MJNeves2018,MJNevesAdP2017}.

In others approaches, Abelian fields with kinetic mixing parameters are also studied as possible effective models of hidden sectors.
Or as a possible Dark Matter (DM) sector, that can connect the barionic matter of SM to an unknown DM content.
In this case, the DM scenario could be observed through the wave interference phenomena, or through the scattering
of known particles, like quarks and electrons, into a DM content. In this scenario, the boson that mixes with the photon
is called dark photon, or hidden photon.

In this work, we will discuss a new issue in the literature, i.e., $X$-boson classical electrodynamics and its gauge symmetry, based on the effective Lagrangian in Eq. (\ref{Lagrangiana}). We have also added a classical source $J^{\mu}$, due to the interaction $J_{\mu}A^{\mu}$, with the $A^{\mu}$-field. We obtain the field equations, and the correspondent wave equations for tensor fields. The diagonalization of Eq. (\ref{Lagrangiana}) is an important tool to understand explicitly that the model is composed by both Maxwell and Proca sectors where the mass is given by $m_{X}/\sqrt{1-\chi^2}$. Furthermore, we believe that this way is the simplest one to obtain the conservation laws and the conserved components of the energy-momentum tensor. Moreover, we obtain the inhomogeneous solution for the potentials
$\{ \, A^{\mu} \, , \, X^{\mu} \, \}$ and for the tensor fields, in the case of a point charge moving in the space.
This is known as Jefimenko's equations for the point charge. Finally, as an application, we will obtain the components
of the $X$-field tensor for the uniform motion of a point charge.

In this paper we will obey the following distribution of the subjects.  All the following results are new in the literature.  In
section II, we have described the first steps of the $X$-boson electrodynamics.  In section III, a gauge invariant action for the $X$-boson was obtained. In section IV, the conserved energy-momentum tensor was discussed. The potentials for the charge motion and the fields were analyzed in section V.  The conclusions were left for the last section.
\section{Maxwell's and wave equations}

Let us start this section with the classical field equations relative to the effective Lagrangian in Eq. (\ref{Lagrangiana}).
The action principle applied to the Lagrangian yields the following field equations\footnote{We are adopting here
the natural units $\hbar=c=1$.} :
\begin{eqnarray}\label{EqsFX}
\partial_{\mu}F^{\mu\nu} + \chi \, \partial_{\mu}X^{\mu\nu} \!&=&\! J^{\nu} \label{Jnu} \; ,
\\
\partial_{\mu}X^{\mu\nu} + \chi \, \partial_{\mu}F^{\mu\nu} + m_X^2 \, X^{\nu} \!&=&\! K^{\nu}\label{Knu} \; ,
\end{eqnarray}
where we have introduced the classical source $J^{\mu}$, due to the coupling $J_{\mu}A^{\mu}$. Furthermore,
the antisymmetrical tensors satisfy the usual Bianchi identities
\begin{eqnarray}\label{bianchiX}
\partial_{\mu}F_{\nu\rho}+\partial_{\nu}F_{\rho\mu}+\partial_{\rho}F_{\mu\nu} \!&=&\! 0
\nonumber \\
\partial_{\mu}X_{\nu\rho}+\partial_{\nu}X_{\rho\mu}+\partial_{\rho}X_{\mu\nu} \!&=&\! 0 \; .
\end{eqnarray}
In Eqs. (\ref{Jnu}) and  \eqref{Knu} we have two coupled differential equations by the $\chi$-parameter, and in the limit
$\chi \rightarrow 0$, the Maxwell sector is decoupled to a Proca sector.
It is easy to verify in Eq. (\ref{EqsFX}) that $J^{\mu}$-current is conserved,
and in Eq. \eqref{Knu}, the conservation of $K^{\mu}$-current imposes the subsidiary condition $\partial_{\mu}X^{\mu}=0$.

The wave equations for antisymmetric tensors are given by
\begin{eqnarray}\label{eqondaF}
(1-\chi^2) \, \Box F_{\mu\nu} - \chi \, m_X^2 \, X_{\mu\nu} \!\!&=&\!\! \partial_{[\mu}J_{\nu]} - \chi \, \partial_{[\mu}K_{\nu]} \; ,
\nonumber \\
\left(\Box + m_X^2\right) \, X_{\mu\nu} + \chi \, \Box F_{\mu\nu} \!\!&=&\!\! \partial_{[\mu}K_{\nu]} \; ,
\end{eqnarray}
where we have defined the four-curl for classical sources $\partial_{[\mu}J_{\nu]}$ and $\partial_{[\mu}K_{\nu]}$.
%
%
We have considered the vacuum propagation in which $J^{\mu}=K^{\mu}=0$, thereby, the wave equations can be
written in the matrix form
\begin{eqnarray}\label{matriz1}
\left(
\begin{array}{cc}
\left(1-\chi^2\right)\, \Box & \quad -\chi \, m_X^2
\\
\\
\chi \, \Box & \quad \Box + m_X^2
\end{array}
\right)
\left(
\begin{array}{c}
F_{\mu\nu}
\\
\\
X_{\mu\nu}
\end{array}
\right) = 0 \; .
\end{eqnarray}
We have used the wave plane solutions to obtain the correspondent frequencies
\begin{eqnarray}
F^{\mu\nu}(x) = F_{0}^{\, \mu\nu} \, e^{ \, i \, k \, \cdot \, x} \; ,
\nonumber \\
X^{\mu\nu}(x) = X_{0}^{\, \mu\nu} \, e^{ \, i \, k \, \cdot \, x} \; ,
\end{eqnarray}
where $F_{0}^{\, \mu\nu}$ and $X_{0}^{\, \mu\nu}$ are the field amplitudes for both photon and $X$-Boson, respectively, and $k^{\mu}=\left(\omega,{\bf k}\right)$. Thus, the matrix equation (\ref{matriz1}) takes the form in the momentum space
\begin{eqnarray}
\left(
\begin{array}{cc}
\left(1-\chi^2\right) \, k^2 & \quad \chi \, m_{X}^2
\\
\\
\chi \, k^2 & \quad k^2-m_X^2
\end{array}
\right)
\left(
\begin{array}{c}
F_{0}^{\, \mu\nu}
\\
\\
X_{0}^{\, \mu\nu}
\end{array}
\right) = 0 \; .
\end{eqnarray}
The frequencies are calculated by the $2 \times 2$ determinant in which we have the dispersion relations
$k^2 = 0$ and  $k^2 = \tilde{m}_X^{2}$, where the $X$-mass is redefined by
$\tilde{m}_{X}:=m_{X}/(1-\chi^2)$. This last result yields the frequencies for the photon, and for the massive wave
\begin{eqnarray}
\omega({\bf k}) = |{\bf k}|
\hspace{0.5cm} \mbox{and} \hspace{0.5cm}
\omega({\bf k}) = \sqrt{{\bf k}^{2} + \tilde{m}_X^2} \; .
\end{eqnarray}
These results indicate that the mixed equations have degrees of freedom relative to the propagation of a massless particle,
and also to a massive particle with a mass given by $\tilde{m}_{X}$, respectively. This fact must be within the effective
Lagrangian.  We can see it through the full diagonalization of the Lagrangian in Eq. (\ref{Lagrangiana}).
The diagonalization must reveal a Proca sector with mass $\tilde{m}_{X}$ decoupled from the Maxwell sector.
We will see it in detail in the section III. But, before it, we will discuss in details the gauge invariance of the corresponding
action in Eq. (\ref{Lagrangiana}).
\section{Gauge invariance}
As we have said before, the X-boson mass term spoils the gauge invariance of the Lagrangian in Eq. \eqref{Lagrangiana}.  We will use the Noether formalism to recover its gauge invariance. Namely, to obtain a dual action that is gauge invariant, the Noether formalism was used recently in the QFT interference phenomena, string theory and issues about braneworld scenarios, duality and Lorentz invariance violation 
\cite{VahidPRD2016,DalmaziPRD2009,DeriglazovPRD2007,DalmaziPRD2006,DalmaziAbreuPRD2009,EAbreuPLB2011,EAbreuPRD2005,BanerjeeNPB98,EAbreu98}.

In few words, it is a method that establishes a gauge transformation and uses auxiliary fields to obtain a new Lagrangian which is invariant under the established gauge transformation. Of course, the auxiliary fields are eliminated through their equations of motion.

Let us begin by analyzing the Lagrangian in Eq. \eqref{Lagrangiana}


\ni where both the fields and the parameters were defined in the first section. Now we have to establish the gauge transformation that must be obeyed by the fields $A^{\mu}$ and $X^{\mu}$ such that,

\ba
\label{B}
\delta A^\mu &=& \p^\mu \epsilon \quad \therefore \quad \delta F^{\mu\nu} = 0 \; ,
\nonumber \\
\delta X^\mu &=& \p^\mu \alpha \quad \therefore \quad \delta X^{\mu\nu} = 0 \; ,
\nonumber \\
\delta K^\mu &=& 0 \; ,
\ea

\ni and, as we have said before, both the mass and the fermionic terms spoil the gauge invariance.  We have also that $\epsilon$ and $\alpha$ are different gauge parameters.  So,  the variation of (\ref{Lagrangiana}) is

\ba
\label{C}
\delta {\cal L} &=& - \, \frac 12 F_{\mu\nu} \delta F^{\mu\nu}\,-\, \frac 12 X_{\mu\nu} \delta X^{\mu\nu}
\nonumber \\
&+& \frac{\chi}{2} \, \Big(X_{\mu\nu} \delta F^{\mu\nu} \,+\, \delta X_{\mu\nu} F^{\mu\nu} \Big)
\nonumber \\
&+& m^2_X \,X_\mu \delta X^\mu\,-\, K_\mu \delta X^\mu \, \, .
\ea

Substituting in \eqref{C} the gauge transformation written in  \eqref{B} we have that

\be
\delta {\cal L} \,=\, m^2_X X_\mu \p^\mu \alpha \,-\, K_\mu \p^\mu \alpha\,\,,
\ee

\ni which, after an integration by parts, it can be written as

\ba
\label{D}
\delta {\cal L} &=& -m_X^2 \alpha\, \p^\mu X_\mu\,+\, \alpha\, \p^\mu K_\mu
=J \, \alpha
\ea

\ni where

\be
\label{E}
J\,=\,-m_X^2\,\p^\mu X_\mu\,+\, \p^\mu K_\mu
\ee

\ni is the Noether current. We have to construct the second Lagrangian of the iterative Noether method such that

\be
{\cal L}_1\,=\,{\cal L} \,-\,J \, B
\ee

\ni and

\be
\label{F}
\delta {\cal L}_1 \,=\, \delta {\cal L}\,-\, B\,(\delta J)\,-\,J (\delta B) \,\,,
\ee

\ni where $B$ is the mentioned auxiliary field that must be eliminated by the equations of motion. From \eqref{E}

\be
\delta J \,=\, -m^2_X\,\p^\mu (\delta X_\mu )\,+\, \p^\mu (\delta K_\mu).
\ee

Now, let us define that $\delta B = \alpha$ and from Eq. \eqref{F} we have that

\ba
\label{G}
\delta {\cal L}_1 &=& -\,(\delta J)\,B
=m^2_X \, \p^\mu (\delta X_\mu)\, B
\nonumber \\
&=&m^2_X\,(\Box\,\alpha ) \,B=m^2_X \, (\Box\,B ) \, \delta \, B \, \, ,
\ea

\ni where we have integrated by parts twice. Finally we can write that

\be
\label{H}
{\cal L}_2\,=\,{\cal L}_1\,-\,\frac 12 m^2_X \,B\,\Box\,B \, ,
\ee

\ni and, using \eqref{G}, we have finally that $\delta {\cal L}_2\,=\,0$.
The gauge invariant action, substituting both ${\cal L}$ and ${\cal L}_1$ into ${\cal L}_2$, is

\ba
\label{I}
{\cal L}_2 \,&=&\, -\dfrac{1}{4} F_{\mu\nu}^2 - \dfrac{1}{4} X_{\mu\nu}^2 + \dfrac{\chi}{2} X_{\mu\nu}F^{\mu\nu} + \dfrac{1}{2} \, m_X^2 \, X_{\mu}^2
\nonumber \\
&-& K_{\mu}X^{\mu} \,-\, J \, B \,-\, \frac 12 \, m^2_X B \, \Box B
\ea

and, finally, to eliminate the $B$-field,

\be
\label{J}
\frac{\delta{\cal L}_2}{\delta B} \,=\, 0
\quad \therefore \quad
B\,=\,\frac{1}{\Box} \Big(\p^\mu X_\mu\,-\,\frac{1}{m^2_X} \, \p^\mu K_\mu \Big) \, ,
\ee

\ni which means that, although we a have a nonlocal action in \eqref{I}, the Lagrangian is gauge invariant.  The better way to eliminate this nonlocallity is through the addition of auxiliary fields.
The analysis of the Lagrangian in Eqs. \eqref{I} and \eqref{J} within the $X$-boson scenario is an ongoing research.

\section{The formal diagonalization and energy-momentum tensor}
Let us now take a more formal path to diagonalize the effective Lagrangian in Eq. \eqref{Lagrangiana}. First of all, it is useful to recognize that the kinetic terms may be written in the field-operator-field form, so that we have
%
%
\begin{equation}\label{Lag}
\mathcal{L} = \frac{1}{2} \, \left(V^{\mu}\right)^{t} \, \Box\theta_{\mu\nu} \, K \, V^{\nu} + \frac{1}{2} \, \left(V^{\mu}\right)^{t} \, \eta_{\mu\nu} \, M^2 \, V^{\nu}-\left(C_{\mu}\right)^t V^{\mu}  \; ,
\end{equation}
where $\theta_{\mu\nu}=\eta_{\mu\nu}-\omega_{\mu\nu}$ and $$\omega_{\mu\nu}=\frac{\partial_{\mu}\partial_{\nu}}{\Box}$$ are the transverse projectors.
Here, we have defined the 2-row vectors $(V^{\mu})^{t} = \left( A^{\mu} \, \, X^{\mu} \right)$ and $(C^{\mu})^{t} = \left( J^{\mu} \, \, K^{\mu} \right)$, and the matrices $K$ and $M^2$ are given by
\begin{equation}
K = \left(
\begin{array}{cc}
1 & \quad -\chi
\\
\\
-\chi & \quad 1 \\
\end{array}
\right)
\hspace{0.2cm} \text{and} \hspace{0.2cm}
M^2 = \left(
\begin{array}{cc}
0 & \quad 0
\\
\\
0 & \quad m_{X}^2 \\
\end{array}
\right). \label{M1}
\end{equation}
We will begin with the diagonalization of $K$ via an orthogonal matrix $R$, so that the $2$-vector has the transformation $\tilde{V}^{\mu}=R \, V^{\mu}$.
The $K_{D}$-diagonal matrix satisfies the relation $K_{D}=R \, K \, R^{t}$.
The solution is that $R$ is an $SO(2)$-matrix with rotation angle of $45^{o}$. The eigenvalues of $K$ are given by $1 \mp \chi$, namely,
$K_{D} = \left( \, 1-\chi \, , \, 1+\chi \, \right)$. Thereby, the mass matrix $M^{2}$ can be written in the basis $\tilde{V}$ as
%
%
%
\begin{eqnarray}
\tilde{M}^2 = R \, M^2 \, R^{t} =
\frac{m_{X}^2}{2}\left( \begin{array}{cc}
1 & \quad 1
\\
\\
1 & \quad 1
\\
\end{array}
\right) \; .
\end{eqnarray}
The Lagrangian in terms of $\tilde{V}$ is given by

\begin{eqnarray}\label{Lag1}
\mathcal{L} &=& \frac{1}{2} \, ( \, \tilde{V}^{\mu} \, )^{t} \, \Box\theta_{\mu\nu} \, K_{D} \, \tilde{V}^{\nu}
+ \frac{1}{2} \, (\tilde{V}^{\mu})^{t} \, \eta_{\mu\nu} \, \tilde{M}^2 \, \tilde{V}^{\nu} \nonumber \\
&-&(\tilde{C}_{\mu})^t \tilde{V}^{\mu} \; ,
\end{eqnarray}
where $\tilde{C}^{\mu}=R \, C^{\mu}$. We have split $K_{D}$ into $\left( K_{D}^{1/2} \right)^{t} \, K_{D}^{1/2}$ in order to eliminate any explicit matrix object in the kinetic sector.
%
%
%
%
%
After some computation, one can quickly find that a possible solution of the $K_{D}^{1/2}$-matrix presented above is
\begin{equation}
K_{D}^{1/2} = \left(
\begin{array}{cc}
\sqrt{1 - \chi} & \quad 0
\\
\\
0 & \quad \sqrt{1 + \chi}
\\
\end{array}
\right), \label{Ksqrt}
\end{equation}
which indeed verifies all the constraints above and has $K_{D}$ as its square. Thus, we can define $\tilde{\tilde{V}}^{\mu} = K_{D}^{1/2} \, \tilde{V}^{\mu}$, so that
\begin{equation}\label{Lag2}
\mathcal{L} = \frac{1}{2} \, (\tilde{\tilde{V}}^{\mu})^{t} \Box\theta_{\mu\nu} \, \tilde{\tilde{V}}^{\nu}
+ \frac{1}{2} \, (\tilde{\tilde{V}}^{\mu})^{t} \eta_{\mu\nu} \, \tilde{\tilde{M}}^2 \, \tilde{\tilde{V}}^{\nu}
- (\tilde{\tilde{C}}^{\mu})^t \tilde{\tilde{V}}^{\mu} \, ,
\end{equation}
where $\tilde{\tilde{M}}^2 = \left( K_{D}^{1/2} \right)^{-1} \, \tilde{M}^2 \, \left( K_{D}^{1/2} \right)^{-1}$
and $\tilde{\tilde{C}}^{\mu}=\left(K_D^{1/2}\right)^{-1} R \, \tilde{J}_{\mu}$.
%
%
Now we can compute $\tilde{\tilde{M}}^2$ explicitly as
\begin{equation}\label{Ksqrt_inv}
\tilde{\tilde{M}}^2 = \frac{m_{X}^2}{2}\left(
\begin{array}{cc}
\frac{1}{1 - \chi} & \quad \frac{1}{\sqrt{1 - \chi^2}}
\\
\\
\frac{1}{\sqrt{1 - \chi^2}} & \quad \frac{1}{1 + \chi} \\
\end{array}
\right) \; .
\end{equation}
Since $\tilde{\tilde{M}}^2$ is symmetric, it can be diagonalized through another $SO(2)$-orthogonal matrix $S$.
Therefore, let us define $\tilde{\tilde{\tilde{V}}}^{\mu} = S \, \tilde{\tilde{V}}^{\mu}$, so that we will end up with a
totally diagonal Lagrangian, namely

\begin{equation}\label{Lag3}
\mathcal{L} = \frac{1}{2} \, (\tilde{\tilde{\tilde{V}}}^{\mu})^{t} \Box\theta_{\mu\nu} \, \tilde{\tilde{\tilde{V}}}^{\nu} + \frac{1}{2} \, (\tilde{\tilde{\tilde{V}}}^{\mu})^{t} \eta_{\mu\nu} \, M_{D}^2 \, \tilde{\tilde{\tilde{V}}}^{\nu}
-(\tilde{\tilde{\tilde{C}}}^{\mu})^t \tilde{\tilde{\tilde{V}}}^{\mu}   \; ,
\end{equation}
where $M_{D}^{2} = S \, \tilde{\tilde{M}}^2 \, S^{t}$ and $\tilde{\tilde{\tilde{C}}}^{\mu}=S \, \left(K_D^{1/2}\right)^{-1} R \, C^{\mu}$. Thus, we can obtain it explicitly by extracting the eigenvalues from $\tilde{\tilde{M}}^2$, 
%
\begin{equation}\label{Mdiag}
M_{D}^2 = \left(
\begin{array}{cc}
0 & \quad 0
\\
\\
0 & \quad \frac{m_{X}^2}{1 - \chi^2} \\
\end{array}
\right) \; .
\end{equation}
Notice that we could have included the $\chi$-dependent denominator into the definition of $m^2$ without loss of generality. This shows explicitly that we are in fact dealing with two vector fields, one massless, to be interpreted as the usual Maxwell photon, and the other, a massive $X$-boson. The orthogonal $S$-matrix is written in terms of the $\chi$-parameter
\begin{eqnarray}
S=\frac{1}{\sqrt{2}}
\left(
\begin{array}{cc}
\sqrt{1-\chi} & \quad -\sqrt{1+\chi}
\\
\\
\sqrt{1+\chi} & \quad \sqrt{1-\chi}
\\
\end{array}
\right) \; .
\end{eqnarray}
The Lagrangian in Eq. (\ref{Lag3}) is now composed by a Maxwell and a Proca-type theory where the mass term is given in Eq. \eqref{Mdiag}
so we can write it as
\begin{eqnarray}
{\cal L}=-\frac{1}{4} \tilde{\tilde{\tilde{F}}}_{\mu\nu}^{\, 2}
-\frac{1}{4} \tilde{\tilde{\tilde{X}}}_{\mu\nu}^{\, 2}
+\frac{1}{2} \tilde{m}_{X}^{2} \tilde{\tilde{\tilde{X}}}_{\mu}^{\, 2}
+ \tilde{\tilde{\tilde{J}}}_{\mu} \tilde{\tilde{\tilde{A}}}^{\mu}
+ \tilde{\tilde{\tilde{K}}}_{\mu} \tilde{\tilde{\tilde{X}}}^{\mu}  \; ,
\end{eqnarray}
where the new fields $\{ \, \tilde{\tilde{\tilde{A}}}^{\mu}
\, , \, \tilde{\tilde{\tilde{X}}}^{\mu} \, \}$ can be obtained in terms of the old basis $\{ \, A^{\mu} \, , \, X^{\mu}  \, \}$
through the transformation $\tilde{\tilde{\tilde{V}}}^{\mu}=U \, V^{\mu}$, in which $U$ is the matrix
$U=S \, K_{D}^{\, 1/2} \, R$.
%
%
%
More explicitly, these transformations yield the transformations of the fields and currents
\begin{eqnarray}\label{transfAX}
\tilde{\tilde{\tilde{A}}}^{\mu} \!\!&=&\!\! A^{\mu} - \chi \, X^{\mu}
\quad , \quad
\tilde{\tilde{\tilde{X}}}^{\mu} = \sqrt{1-\chi^{2}} \, X^{\mu}  \; ,
\nonumber \\
\tilde{\tilde{\tilde{J}}}^{\mu} \!\!&=&\!\! J^{\mu}
\quad , \quad
\tilde{\tilde{\tilde{K}}}^{\mu} = \frac{\chi \, J^{\mu}}{\sqrt{1-\chi^{2}}} +\frac{K^{\mu}}{\sqrt{1-\chi^{2}}} \, ,
\end{eqnarray}
and the inverse transformation are
\begin{eqnarray}\label{transfJK}
A^{\mu} \!\!&=&\!\! \tilde{\tilde{\tilde{A}}}^{\mu} + \frac{\chi \, \tilde{\tilde{\tilde{X}}}^{\mu}}{\sqrt{1-\chi^{2}}}
\quad , \quad
X^{\mu}=\frac{\tilde{\tilde{\tilde{X}}}^{\mu}}{\sqrt{1-\chi^{2}}} \; ,
\nonumber \\
J^{\mu} \!\!\!&=&\!\!\! \tilde{\tilde{\tilde{J}}}^{\mu}
\quad , \quad
K^{\mu}= - \dfrac{\chi \, \tilde{\tilde{\tilde{J}}}^{\mu}}{\sqrt{1-\chi^{2}}} + \dfrac{\tilde{\tilde{\tilde{K}}}^{\mu}}{\sqrt{1-\chi^{2}}} \; .
\end{eqnarray}

The field equations are simpler since the mixed terms has been eliminated
\begin{eqnarray}\label{EqsAXtil3}
\partial_{\mu}\tilde{\tilde{\tilde{F}}}^{\mu\nu}=\tilde{\tilde{\tilde{J}}}^{\nu}
\hspace{0.3cm} \mbox{and} \hspace{0.3cm}
\partial_{\mu}\tilde{\tilde{\tilde{X}}}^{\mu\nu}+ m_{X}^{2} \, \tilde{\tilde{\tilde{X}}}^{\nu}= \tilde{\tilde{\tilde{K}}}^{\nu} \; .
\end{eqnarray}
Thus, we have obtained a kind of Maxwell's equation in the presence of classical sources and a Proca's equation with the $X$-mass.
Obviously, the strength field tensors satisfies the Bianchi identities, and now it is easier to work with the basis $\left\{ \, \tilde{\tilde{\tilde{A}}}^{\mu} \, , \, \tilde{\tilde{\tilde{X}}}^{\mu} \, \right\}$.
For example, we have in this stage the energy-momentum tensors for $J^{\mu}=K^{\mu}=0$
\begin{eqnarray}\label{TensorsEne}
\Sigma^{\mu}_{\; \; \, \rho} \!\!&=&\!\! \tilde{\tilde{\tilde{F}}}^{\mu\nu}\tilde{\tilde{\tilde{F}}}_{\nu\rho}
-\delta^{\mu}_{\; \; \, \rho} \left( -\frac{1}{4} \, \tilde{\tilde{\tilde{F}}}_{\alpha\beta}^{2} \right) \, ,
\nonumber \\
\Xi^{\mu}_{\; \; \, \rho} \!\!&=&\!\! \tilde{\tilde{\tilde{X}}}^{\mu\nu}\tilde{\tilde{\tilde{X}}}_{\nu\rho}
+\tilde{m}_{X}^{2} \, \tilde{\tilde{\tilde{X}}}^{\mu} \tilde{\tilde{\tilde{X}}}_{\rho}
\nonumber \\
&&
\hspace{-0.5cm}
-\delta^{\mu}_{\; \; \, \rho} \left( -\frac{1}{4} \, \tilde{\tilde{\tilde{X}}}_{\alpha\beta}^{2}+\frac{1}{2} \, \tilde{m}_{X}^{2} \, \tilde{\tilde{\tilde{X}}}_{\alpha}^{2} \right) \; ,
\end{eqnarray}
and by making the transformation $\left\{ \, \tilde{\tilde{\tilde{A}}}^{\mu} \, , \, \tilde{\tilde{\tilde{X}}}^{\mu} \, \right\}
\mapsto \left\{ \, A^{\mu} \, , \, X^{\mu} \, \right\}$, we obtain Eq. (\ref{TensorsEne}) in the old basis
\begin{eqnarray}\label{TensorsEneBasisAX}
\Sigma^{\mu}_{\; \; \, \rho} \!\!&=&\!\! F^{\mu\nu}F_{\nu\rho}
-\chi \, \left( \, F^{\mu\nu}X_{\nu\rho}+X^{\mu\nu}F_{\nu\rho} \, \right)
+\chi^{2} \, X^{\mu\nu}X_{\nu\rho}
\nonumber \\
&&
\hspace{-0.5cm}
-\delta^{\mu}_{\; \; \, \rho} \left( -\frac{1}{4} \, F_{\alpha\beta}^{2}+\frac{\chi}{2} \, F_{\alpha\beta}X^{\alpha\beta}
-\frac{\chi^{2}}{4} \, X_{\alpha\beta}^{2} \right) \, ,
\nonumber \\
\Xi^{\mu}_{\; \; \, \rho} \!\!&=&\!\! (1-\chi^{2}) \, X^{\mu\nu}X_{\nu\rho}
+m_{X}^{2} \, X^{\mu} X_{\rho}
\nonumber \\
&&
\hspace{-0.5cm}
-\delta^{\mu}_{\; \; \, \rho} \left[ -\frac{1}{4} \left(1-\chi^{2}\right) X_{\alpha\beta}^{2}+\frac{1}{2} \, m_{X}^{2} \, X_{\alpha}^{2} \right] \; .
\end{eqnarray}
Both tensors are symmetric, and satisfy the conservation laws $\partial_{\mu}\Sigma^{\mu}_{\; \; \, \rho}=\partial_{\mu}\Xi^{\mu}_{\; \; \, \rho}=0$.
We can observe that from the results in Eq. (\ref{TensorsEneBasisAX}), some parts of the Lagrangian (\ref{Lagrangiana}) emerge.   Moreover,
we define the total energy-momentum tensor of the model as the sum $\Theta^{\mu}_{\; \; \, \rho}=\Sigma^{\mu}_{\; \; \, \rho}+\Xi^{\mu}_{\; \; \, \rho}$

\begin{eqnarray}
\Theta^{\mu}_{\; \; \, \rho} \!\!&=&\!\! F^{\mu\nu}F_{\nu\rho}-\chi \, \left( \, F^{\mu\nu}X_{\nu\rho}+X^{\mu\nu}F_{\nu\rho} \, \right)
\nonumber \\
&&
\hspace{-0.5cm}
+X^{\mu\nu}X_{\nu\rho}+m_{X}^{2} \, X^{\mu}X_{\rho}- \delta^{\mu}_{\, \, \rho} \, {\cal L} \; ,
\end{eqnarray}
where ${\cal L}$ is the Lagrangian (\ref{Lagrangiana}), when $K^{\mu}=0$. Clearly, the full energy-momentum tensor
$\Theta^{\mu}_{\; \; \, \rho}$, which is also symmetric.   It also satisfies the conservation law.
The conserved components are $\Theta^{0}_{\; \; \, \rho}=\left(\Theta^{0}_{\; \; \, 0} \, , \, \Theta^{0}_{\; \; \, i}\right)$,
where the energy density $\Theta^{0}_{\; \; \, 0}$ of the model is
\begin{eqnarray}
u \!&=&\!
\frac{1}{2}\left( {\bf E}^{2}+{\bf B}^{2} \right) +
\frac{1}{2}\left( {\bf G}^{2}+{\bf H}^{2} \right)
\nonumber \\
&&
\hspace{-0.5cm}
- \, \chi \, \left( {\bf E}\cdot{\bf G}+{\bf B}\cdot{\bf H} \right)+m_{X}^{2} \, \Phi^{\, 2} \; ,
\end{eqnarray}
and the Poynting vector is the $\Theta^{o}_{ \; i}$ component
\begin{equation}
{\bf S}={\bf E} \times {\bf B} + {\bf G} \times {\bf H} - \chi \left( \, {\bf E} \times {\bf H}+ {\bf G} \times {\bf B} \, \right) + m_{X}^{2} \Phi \, {\bf X} \; .
\end{equation}
Here, we can write explicitly the components of the strength field tensors as the electric and magnetic fields, {\it i. e.},
$F^{\mu\nu}=\left(E^{i} \, , \, \epsilon^{ijk}B^{k} \right)$ and the $X$-fields as $X^{\mu\nu}=\left(G^{i} \, , \, \epsilon^{ijk}H^{k} \right)$.
The four-potential components can be defined by $A^{\mu}=\left( \, V \, , \, {\bf A} \, \right)$ and $X^{\mu}=\left( \, \Phi \, , \, {\bf X} \, \right)$.
Let us call $G$ and $H$ as the $X$-fields.

Both the energy and momentum of the model depend both on the $\chi$-parameter which yields the mixed contribution of the original Lagrangian such that, if we make $\chi \rightarrow 0$, we can obtain the Maxwell electromagnetism uncoupled of the Proca model. Since we know the transformations for the diagonal basis, it is easier to obtain the solution, and then, we can return to the old basis by means of the inverse transformation in Eq. (\ref{transfJK}).
\section{Potentials and fields for point charge motion}
Let us go back to the field equations in Eq. (\ref{EqsAXtil3}) to obtain the solution for the $4$-potentials in the basis $\left\{ \, \tilde{\tilde{\tilde{A}}}^{\mu} \, , \, \tilde{\tilde{\tilde{X}}}^{\mu} \right\}$. Thus, this is the simplest way to obtain the solution for the $4$-potentials in the old basis $\left\{ \, A^{\mu} \, , \, X^{\mu} \right\}$. The subsidiary condition $\partial_{\mu}\tilde{\tilde{\tilde{X}}}^{\mu}=0$ is the consequence of the conserved current in Eq. (\ref{EqsAXtil3}), and if we choose the Lorenz gauge $\partial_{\mu}\tilde{\tilde{\tilde{A}}}^{\mu}=0$, the $4$-potential equations will be reduced to the usual form
\begin{eqnarray}\label{EqsAXtil3cond}
\Box \, \tilde{\tilde{\tilde{A}}}^{\mu}=\tilde{\tilde{\tilde{J}}}^{\mu}
\hspace{0.3cm} \mbox{and} \hspace{0.3cm}
\left( \, \Box+\tilde{m}_{X}^{2} \, \right)\tilde{\tilde{\tilde{X}}}^{\mu}=\tilde{\tilde{\tilde{K}}}^{\mu} \; .
\end{eqnarray}
It can be solved by the usual Green function method via Fourier integrals. We write the solutions of Eq.  (\ref{EqsAXtil3cond}) through integrals
\begin{eqnarray}
\tilde{\tilde{\tilde{A}}}^{\mu}\left(x\right) \!\!&=&\!\! \int_{{\cal R}} d^{4}x' \, G^{(-)}(x-x^{\prime}) \, \tilde{\tilde{\tilde{J}}}^{\mu}(x^{\prime}) \; .
\nonumber \\
\tilde{\tilde{\tilde{X}}}^{\mu}\left(x\right) \!\!&=&\!\! \int_{{\cal R}} d^{4}x' \, \Delta^{(-)}(x-x^{\prime}) \, \tilde{\tilde{\tilde{K}}}^{\mu}(x^{\prime}) \; ,
\end{eqnarray}
where $G^{(-)}$ is the retarded Green function of the first equation of (\ref{EqsAXtil3cond}),
and $\Delta^{(-)}$ is the retarded Green function for a massive particle of its second equation.
We use the known results from field theory handbooks for these Green functions, i. e.,
\begin{eqnarray}\label{Greenfunctions}
G^{(-)}(x-x^{\prime}) \!\!&=&\!\! \frac{1}{2\pi} \, \delta\left[(x-x^{\prime})^2\right]
\, \, \mbox{sgn}(\tau)
\nonumber \\
\Delta^{(-)}(x-x^{\prime}) \!\!&=&\!\! \frac{1}{2\pi} \, \delta\left[(x-x^{\prime})^2\right]
\, \, \mbox{sgn}(\tau)
\nonumber \\
&&
\hspace{-2cm}
+ \frac{m_{X}}{4\pi} \, \frac{\Theta\left((x-x^{\prime})^{2}\right)}{\sqrt{(x-x^{\prime})^2}} \, J_{1}\left(m_{X}\sqrt{(x-x^{\prime})^{2}}\right)
\, \, \mbox{sgn}(\tau) \; .
\hspace{0.7cm}
\end{eqnarray}
Here, the notation $\mbox{sgn}$ indicates the signal function of $\tau:=t-t^{\prime}>0$, and $J_{1}$ is the $J$-Bessel function.
It is clear that the result for the massive particle includes the propagation inside the light-cone for the condition $(x-x^{\prime})^{2}>0$.
Thereby, the retarded solution for the 4-potentials $\tilde{\tilde{\tilde{A}}}^{\mu}$ and $\tilde{\tilde{\tilde{X}}}^{\mu}$ are,
respectively, given by
\begin{eqnarray}
\tilde{\tilde{\tilde{A}}}^{\mu}\left( \, t \, , \, {\bf r} \, \right)=\frac{1}{4 \, \pi} \, \int_{{\cal R}} \, d^{3}{\bf r}^{\prime} \, \, \,
\frac{\tilde{\tilde{\tilde{J}}}^{\mu}\left( \, t_{r} \, , \, {\bf r}^{\, \prime} \, \right)}{|{\bf r}-{\bf r}^{ \, \prime}|} \; ,
\end{eqnarray}

\ni and
\begin{eqnarray}
\tilde{\tilde{\tilde{X}}}^{\mu}\left( \, t \, , \, {\bf r} \, \right) \!\!&=&\!\!
\frac{1}{4 \, \pi} \, \int_{{\cal R}} \, d^{3}{\bf r}^{\prime} \, \, \,
\frac{\tilde{\tilde{\tilde{K}}}^{\mu}\left( \, t_{r} \, , \, {\bf r}^{\, \prime} \, \right)}{|{\bf r}-{\bf r}^{ \, \prime}|}
\nonumber \\
&&
\hspace{-1cm}
+\frac{m_{X}}{4\pi} \int d^4x\,' \frac{\tilde{\tilde{\tilde{K}}}^{\mu}(x')}{\sqrt{(x-x^{\prime})^2}} \, J_{1}\left(m_{X}\sqrt{(x-x^{\prime})^{2}}\right)  \nonumber \\
&&
\hspace{-1cm}
\times \, \Theta\left(\left(x-x^{\prime}\right)^{2}\right) \, \Theta(t-t^{\prime}) \; ,
\hspace{0.4cm}
\end{eqnarray}
where $t_{r}=t-|{\bf r}-{\bf r}^{\prime}|$ is the retarded time.
Using these solutions into the transformations in Eq. (\ref{transfJK}), the $4$-potentials $\left( \, A^{\mu} \, , \, X^{\mu} \, \right)$
of the old basis are obtained in terms of the original sources $\left( \, J^{\mu} \, , \, K^{\mu} \, \right)$. Thus,
the retarded solutions are
\begin{eqnarray}
A^{\mu}(x) \!\!&=&\!\! \frac{1}{4 \pi} \int_{{\cal R}} d^{3}{\bf r}^{\prime} \, \,
\frac{J^{\mu}\left( \, t_{r} \, , \, {\bf r}^{\, \prime} \, \right)}{|{\bf r}-{\bf r}^{ \, \prime}|}
\nonumber \\
&&
\hspace{-1cm}
+ \frac{m_{X} \, \chi}{4 \pi} \int_{{\cal R}} d^{4}x^{\prime}
\, \frac{K^{\mu}\left(x^{\prime}\right)}{\sqrt{(x-x^{\prime})^{2}}} \, \times
\nonumber \\
&&
\hspace{-1cm}
\times \, J_{1}\left(m_{X}\sqrt{(x-x')^{2}}\right) \Theta\left(\left(x-x^{\prime}\right)^{2}\right) \, \Theta(t-t^{\prime}) \; ,
\hspace{0.5cm}
\end{eqnarray}
and
\begin{eqnarray}
X^{\mu}(x) \!\!&=&\!\! \frac{1}{4 \, \pi} \, \int_{{\cal R}} \, d^{3}{\bf r}^{\prime} \, \, \,
\frac{\chi \, J^{\mu}(t_{r} \, , \, {\bf r}^{\, \prime})+K^{\mu}\left(t_{r} \, , \, {\bf r}^{\, \prime}\right)}{|{\bf r}-{\bf r}^{ \, \prime}|}+
\nonumber \\
&&
\hspace{-0.5cm}
+\frac{m_{X}}{4 \pi} \int d^{4}x^{\prime}
\, \frac{\chi \, J^{\mu}(x^{\prime})+K^{\mu}\left(x^{\prime}\right)}{\sqrt{(x-x^{\prime})^{2}}}
\, \times
\nonumber \\
&&
\hspace{-1cm}
\times \,
J_{1}\left(m_{X}\sqrt{(x-x')^{2}}\right) \Theta\left(\left(x-x^{\prime}\right)^{2}\right) \, \Theta(t-t^{\prime}) \; .
\hspace{0.5cm}
\end{eqnarray}
where we have neglected the $\chi^{2}$-order terms. The $\chi$-mixing yields the contribution for the $A^{\mu}$ retarded potential.
It is easy to see that when $\chi \rightarrow 0$, we have the standard $A^{\mu}$ solution for electrodynamics, and $X^{\mu}$ is the retarded
solution for a massive classical field. The interesting case that we consider here is to make $K^{\mu}=0$ and $J^{\mu}\neq0$,
that is, we have just the $J^{\mu}$ of the usual electromagnetism, and only this current yields a contribution for the $X^{\mu}$-potential
through the $\chi$-parameter. Thereby, we have the following potentials written as
\begin{eqnarray}\label{SolucaoAXK=0}
A^{\mu}(x) \!\!&=&\!\! \frac{1}{4 \pi} \int_{{\cal R}} d^{3}{\bf r}^{\prime} \, \,
\frac{J^{\mu}\left( \, t_{r} \, , \, {\bf r}^{\, \prime} \, \right)}{|{\bf r}-{\bf r}^{ \, \prime}|} \; ,
\nonumber \\
X^{\mu}(x) \!\!&=&\!\!
\frac{\chi}{4\pi} \int_{{\cal R}} d^{3}{\bf r}^{\prime} \, \,
\frac{J^{\mu}\left( \, t_{r} \, , \, {\bf r}^{\, \prime} \, \right)}{|{\bf r}-{\bf r}^{ \, \prime}|}
\nonumber \\
&&
\hspace{-1cm}
+\frac{m_{X} \, \chi }{4 \pi} \int d^{4}x^{\prime}
\, \frac{J^{\mu}(x^{\prime})}{\sqrt{(x-x^{\prime})^{2}}} \, \times
\nonumber \\
&&
\hspace{-1cm}
\times \, J_{1}\left(m_{X}\sqrt{(x-x')^{2}}\right) \Theta\left(\left(x-x^{\prime}\right)^{2}\right) \, \Theta(t-t^{\prime}) \; .
\hspace{0.5cm}
\end{eqnarray}

The electrostatic case of a static point charge at the origin wherein
$J^{\mu}({\bf r}^{\prime})=\left( \, Q \, \delta^{3}\left({\bf r}^{\prime}\right) \, , \, {\bf 0} \, \right)$,
in which the $A^{0}$-potential is the Coulomb static potential, and the static $X^{0}$-potential is given by
\begin{eqnarray}
X^{0}({\bf r})=\frac{\chi \, Q \,}{4 \pi \, r} \left[ \hspace{-0.2cm} \phantom{\frac{1}{2}} 1+ I_{1}\left( m_{X} \, r \right) \right]\; ,
\end{eqnarray}
where $I_{1}$ is a Bessel function of first order. We observe the Coulomb limit when $m_{X} \, \rightarrow \, 0$, and the
$I$-Bessel function yields a finite contribution near the origin, when $m_{X}r \ll 1$.

We will discuss now the case of a moving $Q$ in space-time, such that the $4$-current is determined
by the proper-time integral
\begin{eqnarray}
J^{\mu}(x') \!\!&=&\!\! Q \int d\tau \, V^{\mu}(\tau) \, \, \delta^{(4)}\left[x^{\prime}-\xi(\tau)\right] \; ,
\end{eqnarray}
where $\tau$ and $V^{\mu}$ are the proper-time and the $4$-velocity of $Q$, respectively, and we denote $\xi(\tau)$ as being
the $4$-position of $Q$ in relation to a reference frame. The $A^{\mu}$-potential for a relativistic moving charge
is well known in the literature. Then we are dealing here just with the $X^{\mu}$-potential. In this case, the
$X^{\mu}$-potential in the covariant form is
\begin{eqnarray}
X^{\mu}(x) \!\!&=&\!\! \frac{\chi \, Q}{2\pi} \int_{-\infty}^{\infty} d\tau \, V^{\mu}(\tau)
\int d^{4}x^{\prime}
\nonumber \\
&&
\hspace{-1cm}
\times \, \delta^{(4)}\left[x^{\prime}-\xi(\tau) \right]
\delta^{(4)}\left[ (x-x^{\prime})^{2} \right] \, \Theta(t-t^{\prime})
\nonumber \\
&&
\hspace{-1cm}
+\frac{m_{X} \, \chi \, Q}{4 \pi} \int_{-\infty}^{\infty} d\tau \, V^{\mu}(\tau) \int_{{\cal R}} d^{4}x^{\prime}
 \, \frac{\delta^{(4)}[x^{\prime}-\xi(\tau)]}{\sqrt{(x-x^{\prime})^{2}}} \,
\nonumber \\
&&
\hspace{-1cm}
\times \, J_{1}\left(m_{X}\sqrt{(x-x')^{2}}\right) \Theta\left[ (x-x^{\prime})^{2} \right]\Theta(t-t^{\prime}) \; ,
\end{eqnarray}
The first integral is just the same one from the 4-potential $A^{\mu}$ with charge $\chi \, Q$.   The second integral is evaluated with $d^{4}x^{\prime}$ by changing the integration order with $d\tau$.   The function $\Theta(t-\xi^{0}(\tau))$ turns the integral into
%
\begin{eqnarray}\label{Xintdtau}
X^{\mu}(x) \!\!&=&\!\! \left.
\frac{\chi \, Q}{4\pi} \, \frac{V^{\mu}(\tau)}{\left[ x-\xi(\tau) \right] \cdot V(\tau)} \right|_{\tau=\tau_{r}}
\nonumber \\
&&
\hspace{-1cm}
+ \frac{m_{X} \, \chi \, Q}{4 \pi} \int_{-\infty}^{\tau_{r}} d\tau \, \frac{V^{\mu}(\tau)}{\sqrt{(x-\xi(\tau))^{2}}}
\nonumber \\
&&
\hspace{-1cm}
\times \,
\, J_{1}\left(m_{X}\sqrt{(x-\xi(\tau))^{2}}\right) \, \Theta\left[ (x-\xi(\tau))^{2} \right] \; ,
\end{eqnarray}
where the function $\Theta(t-\xi^{0}(\tau))$ solves the integral in $d\tau$ for the interval $-\infty$ to $\tau_{r}$,  where $\tau_{r}$ is the retarded proper time. The condition $(x-\xi(\tau))^{2}>0$ must be imposed concerning the motion of the particle's space-time trajectory. To simplify the integral, we can change the variables $\tau \, \rightarrow \, z:=m_{X} \, \sqrt{(x-\xi(\tau))^{2}}$, so that
%
\begin{eqnarray}
z \, \frac{dz}{d\tau}=-m_{X}^{2} \, \left[x-\xi(\tau)\right]\cdot V(\tau) \; ,
\end{eqnarray}
and the integral in the dimensionlessa variable $z$ turns into
%
\begin{eqnarray}
X^{\mu}(x) \!\!&=&\!\! \left.
\frac{\chi \, Q}{4\pi} \, \frac{V^{\mu}(\tau)}{\left[ x-\xi(\tau) \right] \cdot V(\tau)} \right|_{\tau=\tau_{r}}
\nonumber \\
&&
\hspace{-1cm}
+ \frac{\chi \, Q}{4 \pi} \int_{0}^{\infty} dz \, \frac{V^{\mu}(\tau)}{\left[ x-\xi(\tau) \right] \cdot V(\tau)}
\, J_{1}\left(z\right) \; .
\end{eqnarray}
%
The Bessel function $J_1$ can be reduced to $J_0$ by using the identity $J_{1}(x)=-dJ_{0}/dx$ and an integration by parts.  The derivative term is zero evaluated in $z=0$ and  $z\rightarrow \infty$.  The function $\Theta\left[ (x-\xi(\tau))^{2} \right]$ in Eq. (\ref{Xintdtau}) is equal to one when it is written with the $z$ variable and, recovering the old variable $\tau$, the integral is given by
%
\begin{eqnarray}\label{XmuSolfinal}
X^{\mu}(x) \!\!&=&\!\! \left.
\frac{\chi \, Q}{4\pi} \, \frac{V^{\mu}(\tau)}{\left[ x-\xi(\tau) \right] \cdot V(\tau)} \right|_{\tau=\tau_{r}}+
\nonumber \\
&&
\hspace{-1.5cm}
+ \frac{\chi \, Q}{4 \pi} \int_{-\infty}^{\tau_{r}} d\tau \, \frac{J_{0}\left(m_{X}\sqrt{(x-\xi(\tau))^{2}}\right)}{\left[ x-\xi(\tau) \right] \cdot V(\tau)} \, \times
\nonumber \\
&&
\hspace{-1.5cm}
\times
\left\{ -A^{\mu}(\tau)+ V^{\mu}(\tau) \, \frac{-1+\left[x-\xi(\tau)\right]\cdot A(\tau)}{\left[ x-\xi(\tau) \right] \cdot V(\tau)} \right\} \; ,
\hspace{0.5cm}
\end{eqnarray}
which is the $X$-potential concerning a general point of the space-time system.
To obtain the four-potential $X^{\mu}$ we need to know the motion of the particle through the determination of both its four-velocity and four-acceleration.  It is important to mention that this integral is the contribution of the particle wordline concerning all the motion in the past until the boundary of the lightcone. The fields associated with the $X$-boson are obtained by the computation of $X^{\mu\nu}=\partial^{\mu}X^{\nu}-\partial^{\nu}X^{\mu}$, which is different from the expression in Eq. (\ref{XmuSolfinal}).   However, to obtain an expression for $X^{\mu\nu}$ relative to the $d\tau$ integral is not in the scope of this paper although we can analyze some particular cases for the particle motion.

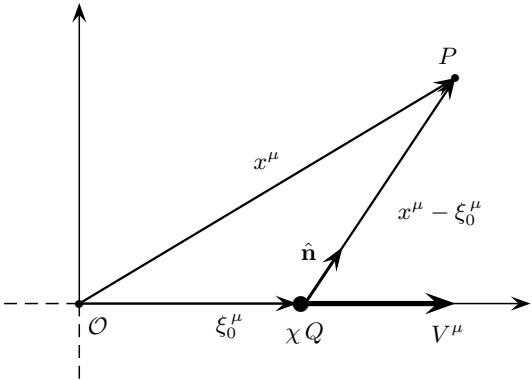
\begin{figure}[h!]
\begin{center}
\newpsobject{showgrid}{psgrid}{subgriddiv=1,griddots=10,gridlabels=6pt}
\psset{unit=1}
\begin{pspicture}(0,2)(7,7)
\psset{arrowsize=0.15 2}
%
%
\rput(1.05,3){\circle*{0.1}}
\rput(1.25,2.7){${\cal O}$}
\psline[linewidth=.2mm,linestyle=dashed](0,3)(1,3)
\psline[linewidth=.2mm]{->}(1,3)(7,3)
\psline[linewidth=.2mm,linestyle=dashed](1,2)(1,3)
\psline[linewidth=.2mm]{->}(1,3)(1,7)
%
%
%
\psline[linewidth=.7mm]{->}(4,3)(6,3)
%
%
%
\rput(4,2.6){$\chi \, Q$}
\rput(4.05,3){\circle*{0.2}}
\rput(5.9,2.6){$V^{\mu}$}
%
%
\psline[linewidth=.3mm]{->}(4,3)(6,6)
\psline[linewidth=.3mm]{->}(1,3)(6,6)
\psline[linewidth=.3mm]{->}(1,3)(3.9,3)
\psline[linewidth=.4mm]{->}(4,3)(4.5,3.75)
\rput(4.05,3.7){\small$\hat{{\bf n}}$}
\rput(5.9,6.3){$P$}
\rput(6.05,6){\circle*{0.1}}
\rput(3,2.7){\small$\xi_{0}^{\, \mu}$}
\rput(5.8,4.2){\small$x^{\mu}-\xi_{0}^{\, \mu}$}
\rput(3.5,4.9){\small$x^{\mu}$}
%
%
%
%
%
%
%
%
%
%
%
\end{pspicture}
\caption{Uniform motion of a point charge $\chi \, Q$ through the horizontal axis}.\label{MovU}
\end{center}
\end{figure}

As an example, one simple case would be the uniform motion of the point charge represented by the figure \ref{MovU}.
The particle 4-position runs with the proper time according to $\xi^{\mu}(\tau)=\xi_{0}^{\mu}+V^{\mu} \tau$,
and the potential $X^{\mu}$ can be written as
%
\begin{eqnarray}
X^{\mu} \!\!&=&\!\! \left. \frac{\chi \, Q}{4\pi} \, \frac{V^{\mu}(\tau)}{\left[ x-\xi(\tau) \right] \cdot V(\tau)} \right|_{\tau=\tau_{r}}
\nonumber \\
&&
\hspace{-1cm}
-\frac{\chi \, Q}{4\pi} \, V^{\mu} \, \int_{-\infty}^{\tau_{r}} d\tau \, \frac{J_{0}\left(m_{X}\sqrt{(x-\xi_{0}-V\tau)^{2}}\right)}
{(\left[ x-\xi_{0}-V \tau \right] \cdot V)^{2}} \; .
\end{eqnarray}
%
%

\noindent
We have considered the position four-vector $x^{\mu}$ and $\xi^{\mu}(\tau)$ is defined by $t-\xi^{0}(\tau_{r})=|{\bf x}-{\bf r}(\tau_{r})|=R$ as the result of the constraint $(x-\xi(\tau_{r}))^{2}=0$.   So, the scalar four-product $V\cdot [x-\xi(\tau_{r})]$ considered in $\tau_{r}$ is
%
\begin{eqnarray}
V.[x-\xi(\tau_{r})]=\gamma \, R - \gamma \, R \, {\bf v}\cdot\hat{{\bf n}}=\gamma \, R \, \left(1-{\bf v}\cdot\hat{{\bf n}}\right) \; ,
\end{eqnarray}
where we have used the four-velocity $V^{\mu}=\left( \, \gamma \, , \, \gamma \, {\bf v} \, \right)$, the position four-vector is
 $(x-\xi)^{\mu}=\left(R,R \, \hat{{\bf n}}\right)$, and $\gamma=\left(1-{\bf v}^2\right)^{-1/2}$ is the Lorentz factor.   Besides, the four-acceleration is $A^{\mu}=\left( \, \gamma^{4} \, {\bf v}\cdot{\bf a} \, , \, \gamma^{2} \, {\bf v}+\gamma^{4} \, {\bf v} \left( {\bf v} \cdot {\bf a}\right) \, \right)$.

To solve the integral in $\tau$, we have to make the variable transformation  $\tau \, \Longrightarrow \, z:=m_{X}\sqrt{(x-\xi_{0}-V \, \tau)^{2}}$, where $z \, dz=-m_{X}^{2}(x-\xi_{0}-V \tau)\cdot V \, d\tau$, and the integral, now in $dz$ can be written as
\begin{eqnarray}
X^{\mu}=\frac{\chi \, Q}{4\pi} \frac{V^{\mu}}{\gamma \, R \left(1-\hat{{\bf n}} \cdot {\bf v} \right)}
+\frac{\chi \, Q \, m_{X}}{4\pi} \, V^{\mu} \, \times
\nonumber \\
\times \int_{0}^{\infty} dz \, \frac{z \, J_{0}\left(z\right) }{\left[z^2+m_{X}^{2}(\left[(x-\xi_{0}) \cdot V\right]^{2}-\left(x-\xi_{0}\right)^2)\right]^{3/2}} \; .
\end{eqnarray}
The $z$ integral is well known in the literature \cite{GradshteynRyzhik00}, i. e.,
\begin{eqnarray}
\int_{0}^{\infty} dz \, \frac{z \, J_{0}(z) }{(z^2+a^2)^{3/2}}=\frac{e^{-\sqrt{a^2}}}{\sqrt{a^2}}
\quad \mbox{if} \quad \Re(a^2)>0 \; .
\end{eqnarray}
Hence, the potential $X^{\mu}$ concerning the uniform motion is
\begin{eqnarray}\label{XmuMU}
X^{\mu}(x)&=&\frac{\chi \, Q}{4\pi} \frac{V^{\mu}}{\gamma \, R \left(1-\hat{{\bf n}} \cdot {\bf v} \right)} \nonumber \\
&+&\frac{\chi \, Q}{4\pi} \, \frac{V^{\mu}}{\sqrt{\left[(x-\xi_{0}) \cdot V\right]^{2}-(x-\xi_{0})^2}} \nonumber \\
&\times& e^{-m_{X}\sqrt{[(x-\xi_{0}) \cdot V]^{2}-(x-\xi_{0})^2}} \; ,
\end{eqnarray}
where $[(x-\xi_{0})\cdot V]^{2}>(x-\xi_{0})^2$.  In fact, by using the definitions, we have that
%
\begin{eqnarray}
[(x-\xi_{0}) \cdot V]^{2}-(x-\xi_{0})^2 \! &=& \! R^{2}+\gamma^{2}R^{2}\left(\hat{{\bf n}}\cdot{\bf v}\right)^{2}
\nonumber \\
\! &=& \! \gamma^{2}R^{2}\left[1-\left({\bf n}\times{\bf v}\right)^{2}\right] ,
\hspace{0.2cm}
\end{eqnarray}
and the potential in Eq. (\ref{XmuMU}) can be written as
%
\begin{eqnarray}\label{XmuMUresult}
X^{\mu}&=&\frac{\chi \, Q}{4\pi} \frac{V^{\mu}}{\gamma \, R \left(1-\hat{{\bf n}} \cdot {\bf v} \right)}
\nonumber \\
&+&\frac{\chi \, Q}{4\pi} \, \frac{V^{\mu}}{\gamma \, R} \, \frac{e^{-\gamma \, m_{X} \, R\sqrt{ 1-(\hat{{\bf n}} \times {\bf v} )^{2}}}}{\sqrt{1- (\hat{{\bf n}} \times {\bf v})^{2} }} \; .
\end{eqnarray}
The tensor $X^{\mu\nu}$ appear from the curl operation $X^{\mu\nu}=\partial^{\mu}X^{\nu}-\partial^{\nu}X^{\mu}$, so the derivative related to $x^{\mu}$ yields
%
\begin{eqnarray}\label{XmunuMU}
X^{\mu\nu} \!\!&=&\!\!
\frac{\chi \, Q}{4\pi} \, \frac{\left[x^{\mu}-\xi^{\mu}(\tau)\right] V^{\nu} - \left[x^{\nu}-\xi^{\nu}(\tau)\right] V^{\mu}}{\left([x-\xi(\tau)]\cdot V\right)^{3}}+
\nonumber \\
&&
\hspace{-0.5cm}
+ \frac{\chi \, Q}{4\pi} \, \left[ \left(x-\xi_{0}\right)^{\mu}V^{\nu}
-\left(x-\xi_{0}\right)^{\nu}V^{\mu} \right] \, \times
\nonumber \\
&&
\hspace{-0.5cm}
\times \, \left( \frac{1+m_{X} \sqrt{[(x-\xi_{0}) \cdot V]^{2}-\left(x-\xi_{0}\right)^{2}} }{\left([(x-\xi_{0}) \cdot V]^{2}-\left(x-\xi_{0}\right)^{2}\right)^{3/2}} \right) \nonumber \\
&\times& e^{-m_{X}\sqrt{[(x-\xi_{0}) \cdot V]^{2}-\left(x-\xi_{0}\right)^{2}}} \; .
\end{eqnarray}
The components of the tensor $X^{\mu\nu}$ can defined by $X^{\mu\nu}=\left( \, G^{i} \, , \, \epsilon^{ijk}H^{k} \, \right)$, in which the vector representing the field ${\bf G}$ from Eq. (\ref{XmunuMU}) is
%
\begin{eqnarray}
{\bf G} \!\!&=&\!\! \frac{\chi \, Q}{4\pi} \frac{\hat{{\bf n}}-{\bf v}}{\gamma^{2}R^2(1-{\bf n}\cdot{\bf v})^{3}}
 \\
&&
\hspace{-0.5cm}
+\frac{\chi \, Q \, \hat{{\bf n}}}{4\pi} \, \frac{1+m_{X}\gamma R \sqrt{1-\left(\hat{{\bf n}}\times {\bf v}\right)^{2}}}{\gamma^{2}R^{2}\left(1-\left(\hat{{\bf n}}\times {\bf v}\right)^{2}\right)^{3/2}} \, e^{-\gamma \, m_{X} \, R\sqrt{ 1-(\hat{{\bf n}} \times {\bf v} )^{2}}} \; . \nonumber
\end{eqnarray}
The reader can verify that the field ${\bf H}$ is connected to ${\bf G}$ through the vector product ${\bf H}=\hat{{\bf n}} \times {\bf G}$ and so, we have that
%
\begin{eqnarray}
{\bf H}=\frac{\chi \, Q}{4\pi} \frac{{\bf v}\times \hat{{\bf n}}}{\gamma^{2}R^2(1-{\bf n}\cdot{\bf v})^{3}} \; .
\end{eqnarray}
%

%


\section{conclusions and Final remarks}

The very well known and underlying forces of nature that are perceived by the people in everyday life are the gravitational and electromagnetic forces. Both have a high range and are present in the microscopic world. The weak and strong forces are less perceived in our daily routine, since they appear only inside the atomic nucleus. The range of these interactions does not go much farther than a nucleus size.

In this work we have analyzed the main elements of a fifth force, which would be necessary to understand the dark matter content. The main particle in this, let us call, dark matter {\it thread}, would be the so-called $X$-boson, which mixes kineticaly with photon via a spin one Lagrangian.

We have explored the $X$-boson classical electrodynamics and we found the field equation from variational principles. After that, using Bianchi identities, we have provided the respective wave equations and dispersion relations. To decouple the $X$-boson from its field, we have accomplished an effective Lagrangian diagonalization procedure through linear transformations and the structure of $SO(2)$-group.
A Klein-Gordon-kind equation for the $X$-boson field was obtained. We have obtained the energy-momentum tensor, and consequently, the conservation laws to find both the energy density and the Poynting vector. After that, we have used the Green functions method to find
the inhomogeneous solution of the 4-potentials associated with the $X$-boson, given the classical sources, $J^{\mu}$ and
$K^{\mu}$. The interesting case is to make $K^{\mu}=0$ and $J^{\mu}\neq0$. Thereby, we have applied this solution to the
relativistic $Q$-point charge motion. In this particular case, when $K^{\mu}=0$,
we have obtained that the kinetic mixing parameter $\chi$ induces a millicharge $\chi \, Q$, which is the source for both the $4$-potential $X^{\mu}$
and its corresponding field tensor $X^{\mu\nu}$. The simplest case for the $4$-potential $X^{\mu}$ is the uniform motion
of $\chi \, Q$, where we have obtained the $X$-fields for the components of the $X^{\mu\nu}$ vector.


Finally, besides all these new results mentioned above, we have obtained a gauge invariant, nonlocal and dual action to the original $X$-boson one.  To obtain its local version, as well as to find its properties is an ongoing research.

\section*{Acknowledgments}

\ni E.M.C.A.  thanks CNPq (Conselho Nacional de Desenvolvimento Cient\' ifico e Tecnol\'ogico), Brazilian scientific support federal agency, for partial financial support, Grant number 302155/2015-5 and the hospitality of Theoretical Physics Department at Federal University of Rio de Janeiro (UFRJ), where part of this work was carried out.

\end{document}